\documentclass[12pt]{article}

\usepackage{authblk}
\usepackage{graphicx} 
\usepackage{amsmath} 
\usepackage[numbers,sort&compress]{natbib} 
\usepackage{hyperref} 
\usepackage{geometry} 
\usepackage{tabularx}

\title{Generative diffusion model surrogates for mechanistic agent-based biological models}

\author[1,2]{
    Tien Comlekoglu\thanks{Corresponding author. Email: tien.comlekoglu@gmail.com}
}
\author[3,4]{J.~Quetzalcoatl Toledo-Mar\'in}
\author[2]{Douglas W.~DeSimone}
\author[1]{Shayn M.~Peirce}
\author[5]{Geoffrey Fox}
\author[6]{James A.~Glazier}

\affil[1]{Department of Biomedical Engineering, University of Virginia, Charlottesville, VA, USA}
\affil[2]{Department of Cell Biology, University of Virginia, Charlottesville, VA, USA}
\affil[3]{TRIUMF, Vancouver, BC, Canada}
\affil[4]{Perimeter Institute for Theoretical Physics, Waterloo, ON, Canada}
\affil[5]{Biocomplexity Institute and the Department of Computer Science, University of Virginia, Charlottesville, VA, USA}
\affil[6]{Department of Intelligent Systems Engineering, Indiana University, Bloomington, IN, USA}

\date{}

\begin{document}

\maketitle

\begin{abstract}
Mechanistic, multicellular, agent-based models are commonly used to investigate tissue, organ, and organism-scale biology at single-cell resolution. The Cellular-Potts Model (CPM) is a powerful and popular framework for developing and interrogating these models. CPMs become computationally expensive at large space- and time- scales making application and investigation of developed models difficult. Surrogate models may allow for the accelerated evaluation of CPMs of complex biological systems. However, the stochastic nature of these models means each set of parameters may give rise to different model configurations, complicating surrogate model development. In this work, we leverage denoising diffusion probabilistic models to train a generative AI surrogate of a CPM used to investigate \textit{in vitro} vasculogenesis. We describe the use of an image classifier to learn the characteristics that define unique areas of a 2-dimensional parameter space. We then apply this classifier to aid in surrogate model selection and verification. Our CPM model surrogate generates model configurations 20,000 timesteps ahead of a reference configuration and demonstrates approximately a 22x reduction in computational time as compared to native code execution. Our work represents a step towards the implementation of DDPMs to develop digital twins of stochastic biological systems.
\end{abstract}

\section{Introduction}
Multicellular agent-based models are commonly used to investigate and simulate complex biological phenomena. These models require simulating the behavior of many model agents at once to observe emergent phenomena inherent in these biological systems. Each agent often represents individual cells, where each cell responds to other cells and the surrounding simulated environment. Modeling complex biological phenomena at the single-cell level can be computationally expensive. Computationally expensive models may be cumbersome or inconvenient for a user to interrogate or continue to develop upon.

The Cellular-Potts model (CPM) is a stochastic mathematical framework that allows for agent-based modeling of complex biological systems such as cells and tissues \citep{swat2012multi}. Accessible, open-source frameworks that facilitate model creation using the CPM such as Compucell3D \citep{swat2012multi}, Morpheus \citep{starruss2014morpheus}, and Artistoo \citep{wortel2021artistoo} integrate the CPM model with other mathematical methods such as ordinary differential equations or boolean networks for subcellular element modeling or partial differential equations (PDEs) to represent diffusive chemical fields. CPMs using these frameworks have been used to investigate tissue and organ-scale complex biological phenomena such as vasculogenesis \citep{merks2006cell}, skeletal muscle regeneration \citep{haase2024agent}, and embryonic development \citep{comlekoglu2024modeling, berkhout2025computational, adhyapok2021mechanical} at single-cell resolution. These agent-based models are used to investigate the state of a biological system by altering parameters that represent the \textit{in silico} perturbation of biological mechanisms, and thus are described as mechanistic models. These models become computationally expensive at large spatial scales and long timescales \citep{chen2007parallel, wright1997potts, gusatto2005efficient}.

Development of a surrogate model may be an effective approach to accelerate the computational evaluation of these multi-scale, multicellular computational models. Deep neural networks have already been applied as effective model surrogates to solve systems of PDEs for physical systems such as heat transfer \citep{farimani2017deep, edalatifar2021using}, molecular particle dynamics \citep{doerr2021torchmd}, or chemical diffusive fields \citep{toledo-marin2021deep, toledo-marin2023analyzing, li2020reaction, fox2019learning}. While methods of applying deep neural networks as surrogates for deterministic physical systems have been investigated, the development of these surrogates for CPM models has not yet been well described. 

Applying neural networks as surrogates for the CPM presents additional challenges compared to that of deterministic systems of PDEs such as heat transfer. Agent-based models often describe emergent behaviors that arise from the interaction of multiple agents within the model that are not explicitly encoded beforehand. CPM models are stochastic, making deterministic neural-network architectures ill-posed to serve as model surrogates. Despite these challenges, the authors have developed a neural-network based surrogate for a CPM of vasculogenesis, where biological vascular patterns emerge from the actions of individual cell agents reacting to each other and a chemotactic diffusive field governed by a system of PDEs \citep{comlekoglu2025surrogate}. This surrogate predicted model configurations at an intermediate timestep of 100 Monte-Carlo steps for a single set of parameters, and was unable to replicate model configurations at long timescales (see Figure \ref{fig:supps1} and Movie S1) or for multiple areas of a parameter space. Because of the stochastic nature of the CPM model, model configurations at long timescales do not retain memory of a reference configuration, or the configuration many computational timesteps prior to the current configuration. Therefore, two equivalent samples from a given area of a model’s parameter space may have different explicit configurations while demonstrating the same phenotypic features that reflect the emergent behaviors of the model mechanisms. These are additional challenges towards developing a surrogate model for a CPM.

In this work, we address these challenges and develop a deep-learning surrogate of a CPM developed for the investigation of \textit{in vitro} vasculogenesis using a denoising diffusion probabilistic model (DDPM). These models have been classically developed to generate novel images by sampling from a learned distribution of data that represents a specific class of image \citep{ho2020denoising}. 

Generative modeling methods such as the DDPM have shown promise as model surrogates for other inherently stochastic, complex systems such as the calorimeter shower for investigating particle interactions in Large Hadron Collider experiments. Multiple generative modeling methods such as Variational Autoencoders (VAEs), Generative Adversarial networks (GANs), Normalizing Flows, Conditional Flow Matching, DDPMs, and Quantum Computing techniques have been applied as calorimeter shower surrogates with variable success \citep{krause2024calochallenge, lu2024zephyr}. DDPM model approaches appear to yield competitive accuracy to other generative modeling methods as a surrogate model, but may be slower than the other methods \citep{krause2024calochallenge}. Recent work in the field of weather forecasting has extended the DDPM to support data assimilation and model personalization yielding effective diffusion model-based surrogates (DiffDA \citep{huang2024diffda} and CorrDiff \citep{mardani2023residual}). This encourages us to use a DDPM to approach the surrogate modeling of our CPM. 

We anticipate that this method of deep generative modeling may learn the distribution of model configurations unique to distinct areas of a CPM parameter space. Sampling from these learned distributions will likely yield a representative sample CPM configuration at a specified long timescale ahead of a reference configuration for different “classes” representing unique areas of the parameter space. We ensure that generated sample configurations exist within the appropriate parameter space by training a neural-network image classifier to perform this verification.

\section{Methods}

\subsection{Cellular-Potts Agent-based Mechanistic Model}
We selected the previously published model of vasculogenesis by Merks et. al. \citep{merks2006cell} because it was an adequate example of an agent-based model replicating a biological system implemented using the Cellular-Potts modeling method with chemical diffusion described by systems of PDEs. The Cellular-Potts (Glazier-Graner-Hogeweg) agent-based model was re-implemented in the CompuCell3D (CC3D) \citep{swat2012multi} open-source simulation environment version 4.6.0. In the CPM, individual cells are represented as a collection of pixels on a square, two-dimensional lattice 256x256 pixels in dimension. Cells are given properties of predefined volume, contact energy with surrounding cells and medium, and a tendency to chemotax toward high concentrations of a diffusive cytokine gradient. These properties are defined mathematically using an effective energy functional $H$ shown in Equation \ref{eq:energy} below. This effective energy functional is evaluated on a cell-by-cell basis each computational timestep, denoted Monte-Carlo step (MCS) in the CPM.
\begin{equation} \label{eq:energy}
\begin{split}
H = & \sum_{i,j,\text{neighbors}} J_{\tau(\sigma_i), \tau(\sigma_j)} (1 - \delta_{\sigma_i, \sigma_j}) \\
  & + \lambda_{\text{volume}} (V_{\text{cell}} - V_{\text{target}})^2 \\
  & + \lambda_{\text{surface}} (S_{\text{cell}} - S_{\text{target}})^2 \\
  & + \sum_{i,j} - \lambda_{\text{chemotaxis}} 
    \left[ 
    \frac{c(\mathbf{x}_{\text{destination}})}{sc(\mathbf{x}_{\text{destination}}) + 1} 
    - 
    \frac{c(\mathbf{x}_{\text{source}})}{sc(\mathbf{x}_{\text{source}}) + 1} 
    \right]
\end{split}
\end{equation}
Where the first term describes cell contact energy for neighboring cells with a contact coefficient $J$ where $i, j,$ describe neighboring lattice sites, $\sigma_i$ and $\sigma_j$ describe individual model agents occupying site $i$ and $j$ respectively, and $\tau(\sigma)$ denotes the type of cell $\sigma$ in the model. The second term defines a volume constraint $\lambda_{\text{volume}}$ applied to each cell where $V_{\text{cell}}$ represents the current volume of a cell at a given point in the simulation, and $V_{\text{target}}$ is the volume assigned to that cell with stiffness $\lambda_{\text{volume}}$. Similarly, the third term applies a surface area constraint, where $\lambda_{\text{surface}}$ determines the influence of assigned circumference $S_{\text{target}}$. The fourth term defines chemotactic agent motility in response to a diffusive gradient where $\lambda_{\text{chemotaxis}}$ is a constraint or influence of the differences in chemical concentration $c(\mathbf{x}_{\text{destination}})$ and $c(\mathbf{x}_{\text{source}})$ of pixel-copy-destination and pixel-copy-source locations of a cell agent each MCS, and $s$ denotes a saturation constant.

At each MCS or computational timestep, the CPM model creates cell movement by selecting random pairs of neighboring voxels $(y, y')$ and evaluating whether one voxel located at $y$ may copy itself to its neighboring pair at $y'$. This voxel copy attempt, denoted $\sigma(y, t) \to \sigma(y', t)$, occurs with the probability defined by a Boltzmann acceptance function (Equation \ref{eq:boltzmann}) of the change in the effective energy of the system $\Delta H$ previously defined in Equation \ref{eq:energy}.
\begin{equation} \label{eq:boltzmann}
\mathrm{Pr}(\sigma(y, t) \to \sigma(y', t)) = e^{-\max\left(0, \frac{\Delta H}{H'}\right)}
\end{equation}
Diffusive chemical concentration values $c$ at each lattice site are described by Equation \ref{eq:diffusion} below:
\begin{equation} \label{eq:diffusion}
\frac{\partial c}{\partial t} = D \nabla^2 c - kc + \text{secretion}
\end{equation}
where $k$ is the decay constant of the diffusive field concentration $c$, and $D$ is the diffusion constant. The secretion term accounts for diffusive field concentration added at lattice sites associated with the location of a CC3D model agent to model secretion of a cytokine by biological cells. Periodic boundary conditions were applied to the simulation domain for both cell positions subject to the Potts algorithm and diffusive field concentrations described by Equation \ref{eq:diffusion}. 

CPM model parameters were defined to produce visually apparent extension of branches, sprouting of new branches, and shrinking of circular lacunae within the parameter space explored previously by Merks et. al. \citep{merks2006cell}. CPM model parameter values are displayed in Table \ref{tab:params}.

\begin{table}[htbp]
\centering
\caption{CPM parameter table}
\label{tab:params}
\begin{tabularx}{\textwidth}{l X l}
\hline
Parameter             & Parameter Description                        & Parameter Value                                    \\ \hline
$\lambda_{\text{volume}}$ & Influence of volume constraint               & 5                                                  \\
$V_{\text{target}}$     & Number of voxels per cell                    & 50                                                 \\
$\lambda_{\text{surface}}$ & Influence of surface constraint              & 1                                                  \\
$S_{\text{target}}$     & Number of voxels in cell circumference       & 16.8                                               \\
$J_{\text{cell,medium}}$ & Contact energy between cell-medium interface & 8.2 at $t_0$, range: 0-20 for experiments \\
$J_{\text{cell,cell}}$   & Contact energy between cell-cell interface   & 6                                                  \\
$\lambda_{\text{chemotaxis}}$ & Influence of chemotaxis                    & 2000                                               \\
$s$                     & Saturation constant for chemotaxis           & 0.5                                                \\
$k$                     & Decay constant for chemical field            & 0.6 at $t_0$, range: 0.05-0.6 for experiments  \\
$H'$                    & Temperature for the Cellular-Potts algorithm & 8                                                  \\ \hline
\end{tabularx}
\end{table}

\subsection{Cellular-Potts Model simulation and training data generation}

\begin{figure}[htbp]
    \centering
    \includegraphics[width=\textwidth]{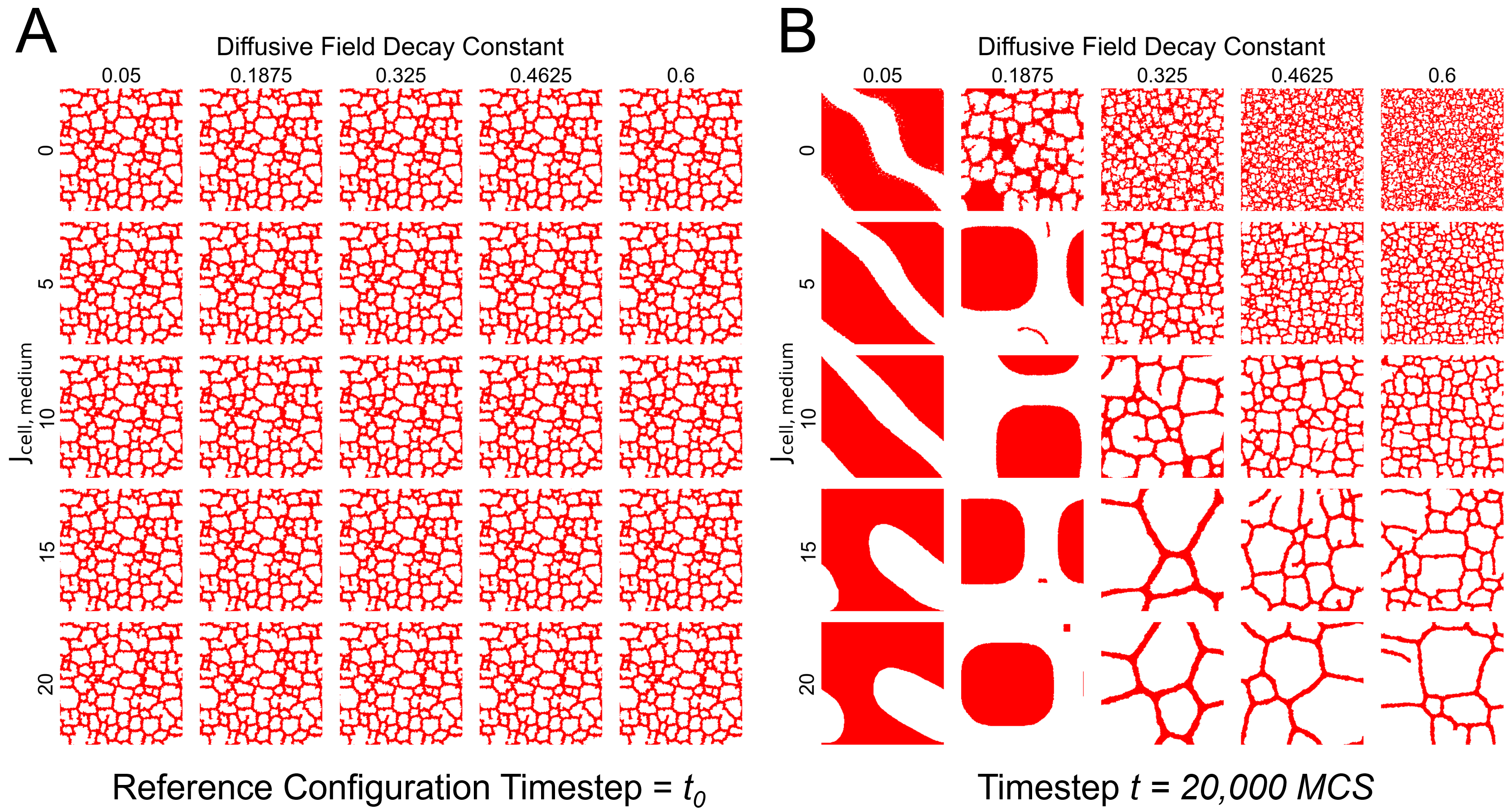}
    \caption{Cellular-Potts Model (CPM) parameters dictate model configuration over long timescales. (A) Reference configuration from a CPM assumes distinct configurations (B) due to changes in mechanistic model parameters over a long timescale of 20,000 simulation timesteps (Monte-Carlo steps, MCS). Memory of the initial configuration is lost but distinct patterns emerge unique to the area of the parameter space.}
    \label{fig:params}
\end{figure}

Approximately 1000 CPM cell agents were placed randomly throughout the 256x256 simulation domain at the beginning of the simulation and settled into vascular-like structures over the first 10,000 CPM simulation timesteps, referred to as Monte-Carlo steps (MCS) of the simulation. This simulation state was saved as an initial timepoint $t_0$ to generate training data for the DDPM surrogate. Training data for the surrogate model was created by saving binary images of the cell layer of model configurations from the CPM model. Images representing model configurations were saved every 100 MCS for a given simulation between 15000 and 20,001 MCS. This procedure was repeated for 100 unique simulations per parameter set spanning a 2-dimensional parameter space. Unique parameter sets were created by defining the parameter $J_{\text{cell,medium}}$ as values within the set $\{0, 5, 10, 15, 20\}$. We also defined the diffusive field decay constant parameter $k$ to be $\{0.05, 0.1875, 0.325, 0.465, 0.6\}$ to define a 5x5 grid of unique parameter sets (Figure \ref{fig:params}A). This sampling of the parameter space was selected to provide a variety of visually distinct phenotypes. While many of these may be less biologically relevant to the developing vasculature that the original CPM was developed to represent, we expected that this variety of visually apparent phenotypes may pose a more challenging dataset for a surrogate to replicate. We also intended for this to result in a method that is generalizable to other computational models, as different visually apparent phenotypes often represent differing system states that would be appropriate to capture with a surrogate model. Over the course of each simulation, the initial configuration $t_0$ with the baseline parameters defined in Table \ref{tab:params} evolved to a new configuration representative of the new parameter set (Figure \ref{fig:params}B, Movie 1). This yielded 5100 images per area of the parameter space and a total of 127500 images for all 25 classes for the total dataset. This dataset was then used to train the diffusion model to serve as a surrogate. This same dataset was also used to train the image classifier for surrogate model verification. A separate dataset was generated in the same manner consisting of 510 images per area of the parameter space for a total of 12750 images used for evaluation of the trained classifier.

\subsection{Denoising Diffusion Model configuration and training}

\begin{figure}[htbp]
    \centering
    \includegraphics[width=\textwidth]{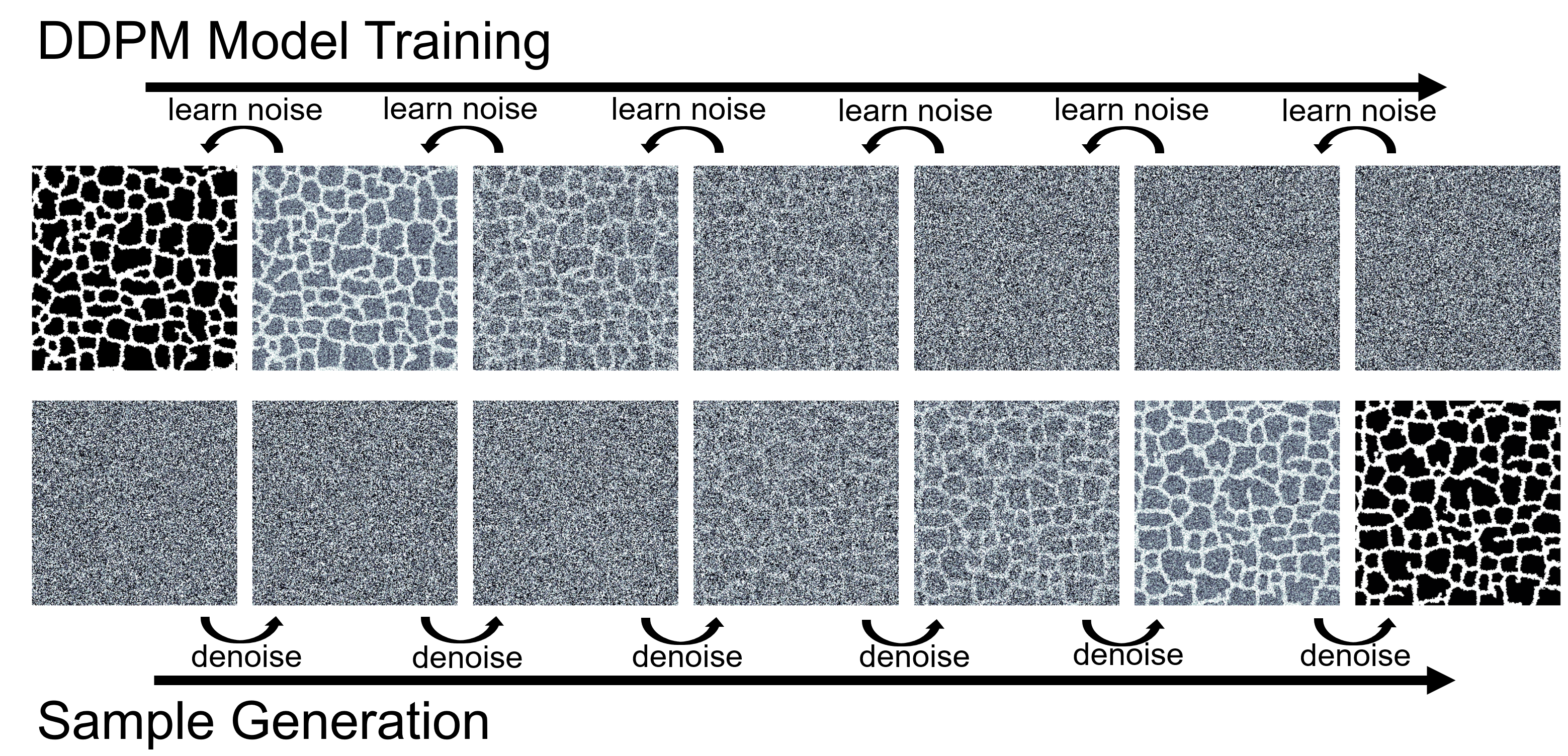}
    \caption{Denoising Diffusion Probabilistic Model (DDPM) training and sample generation. A sample Cellular-Potts model configuration is used for training a DDPM model (top). The trained model may be used to quickly generate a representative model configuration sampled from the same class, or area of the parameter space (bottom).}
    \label{fig:ddpm_train}
\end{figure}

Denoising diffusion probabilistic models (DDPMs) are a generative modeling technique where a convolutional neural network is used to learn a single denoising step of a progressively noisy series of training images. When trained correctly, repeated denoising of an input of gaussian noise reveals an image sampled from the same distribution of data the model was trained with. Figure \ref{fig:ddpm_train} demonstrates the training and generation process for a sample CPM configuration. This results in the generation of an image that could have existed in the original training dataset. Diffusion models may be conditioned on a set of classes during training, thus allowing the model to generate images from various classes of training data \citep{zhan2024conditional}. By defining different areas of the CPM parameter space to be different classes of images, we exploit this functionality to allow our DDPM to generate images from discrete areas of interest in the parameter space of the CPM, therefore allowing it to behave as a surrogate model for the CPM.

To create our CPM surrogate, we used the modified DDPM model architecture, training schema, and fast deterministic sampling method as previously described in the EDM2 method \citep{karras2023analyzing} to create a class-conditional DDPM. We use the edm2-img64-s preset as defined in the original publication but modify the model inputs to allow for single-channel images as well as vary EDM2 training parameters P\_mean and P\_std to define a more optimal set for our use case as is recommended in the original EDM2 publication. P\_mean corresponds to an average amount of gaussian noise to be applied to an image during training of the DDPM, and P\_std corresponds to the variability of noise about P\_mean. We use this model preset to perform diffusion directly in pixel space of our single-channel 256x256 images for a total of 12,748,800 training steps. We chose this setup because at the time of this investigation, the training presets described produce state-of-the-art DDPM model performance for image synthesis for both generation in pixel space and latent space of a pretrained autoencoder for natural images. Thus, we expect these procedures to allow for good performance for our intended use case. Training was performed on 16 V100 GPUs. This preset had the additional benefit of having the smallest computational footprint of the defined presets and thus may be easier to train and more accessible to the broader scientific community.

\subsection{Image Classifier}
To verify that CPM configurations are generated within the appropriate parameter space, we trained and applied a neural network image classifier. We used the EfficientNetV2 architecture because it has been designed to train quickly and retain competitive classification performance as compared to much larger models \citep{tan2021efficientnetv2}. We modified the first layer of the EfficientNetV2-s model to accept single channel images rather than 3-channel RGB images to fit our data and trained the model for 100 epochs. Training data from the CPM was divided into an 80/20 train/validation split and a separate test dataset was used for evaluation of the trained classifier. We used the adaptive moment estimation (Adam) optimizer \citep{kingma2014adam} with a learning rate of 0.001 to achieve an accuracy of 95.6\% on the separate test dataset.

\section{Results}

\subsection{Surrogate Model Selection}

\begin{figure}[htbp]
    \centering
    \includegraphics[width=\textwidth]{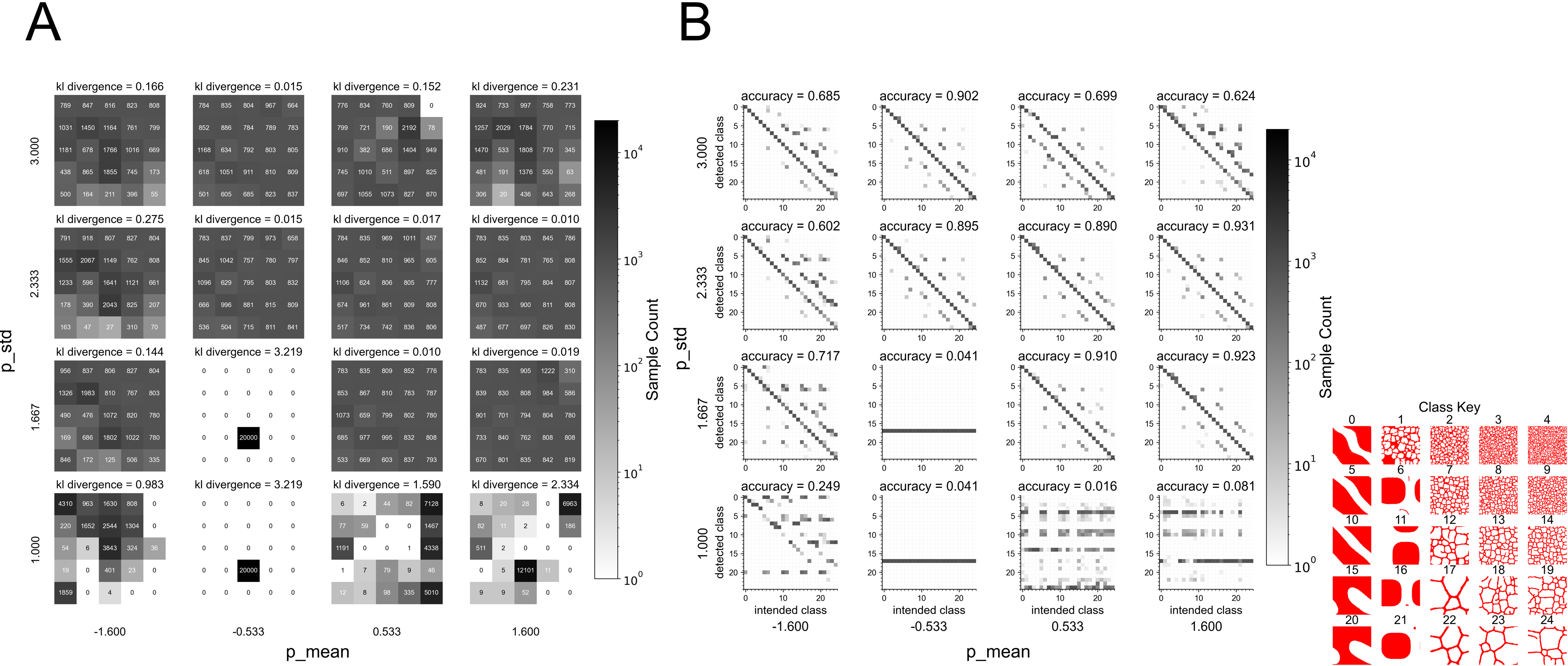}
    \caption{Image classifier identifies mode collapse in trained candidate DDPM surrogate models. DDPM surrogate model performance is sensitive to training hyperparameters P\_mean and P\_std. Each candidate model was used to generate 20,000 configurations uniformly distributed among the 25 classes, or discrete areas of the CPM parameter space. (A) Generated outputs were classified using a trained image classifier and sample counts per class are displayed as a heatmap in the equivalent 5x5 grid layout shown previously in Figure \ref{fig:params}B. Sample counts per class (or discrete parameter set) are displayed within each corresponding cell of each heatmap subplot. KL divergence between generated sample distributions and a uniform distribution is displayed for each DDPM model to provide a quantitative measure of uniformity. KL divergence values close to 0 indicate a more uniform distribution among classes in generated samples. (B) Confusion matrix plots demonstrating classifier identified class (detected class) compared with the conditioning label for generation (intended class) for each model. Generation accuracy is displayed for each DDPM model.}
    \label{fig:collapse}
\end{figure}

We performed a coarse-grained study of training parameters P\_mean and P\_std to attempt to identify a set of parameters that would best allow the DDPM surrogate learn the distributions of model configurations defined in the training set. We defined P\_mean to range from -1.6 to 1.6, and P\_std to range from 1.0-3.0 for a total of 25 candidate models. This parameter range encompassed the parameters found to be optimal for image synthesis in pixel space when trained on natural images as described in the EDM2 work \citep{karras2023analyzing}. For each candidate model, we used the model snapshot saved at the final training step to generate 20,000 CPM configurations uniformly distributed among all 25 parameter spaces (classes). Each sample was generated with a unique random seed (0-19,999 for a total of 20,000 unique samples). We evaluated each model’s ability to generate CPM configurations within all classes by verifying generated outputs with the trained EfficientNetV2-s classifier. Generated class distributions were evaluated for uniformity through calculating the KL divergence between the class-distribution of generated samples and a uniform distribution for each candidate DDPM model. For each trained DDPM surrogate candidate, we plotted the class frequency of CPM configurations generated from the set of 20,000 (Figure \ref{fig:collapse}A). We additionally plot confusion matrices to determine the accuracy of class-conditional generation (intended class) as compared with the sample that was generated as classified by the classifier (detected class) in Figure \ref{fig:collapse}B. Evaluating the candidate surrogate models in this manner revealed that certain combinations of P\_mean and P\_std lead to mode collapse, or a loss in the diversity of output of the DDPM model. Candidate models were expected to produce samples uniformly distributed across the 25 classes because the training data consisted of a uniform distribution among the classes. Of the candidate models, the surrogate trained with P\_mean = 1.6 and P\_std = 1.667 was selected because it generated the most uniform distribution of samples by measure of KL divergence from a uniform distribution, and the greatest class accuracy from class-conditional generation. This model surrogate was then used to generate mechanistic model sample configurations in an unsupervised manner in application as the CPM surrogate model.

We also quantified the vascular morphologic features of branch width and distributions of lacunae area for generated samples. We used Earth Mover's Distance (EMD) to quantify distances between distributions of these morphologic features from DDPM generated configurations and the training data for each class, for each model (Figure \ref{fig:supps2}). Large EMD values indicate a distribution of features for samples of a specific class that deviate significantly from the corresponding training data for that class or area of the CPM parameter space. Absent values indicate samples from which these features could not be quantified, such as samples consisting entirely of background with no generated vessel network.

\subsection{Surrogate Model Application}

\begin{figure}[htbp]
    \centering
    \includegraphics[width=\textwidth]{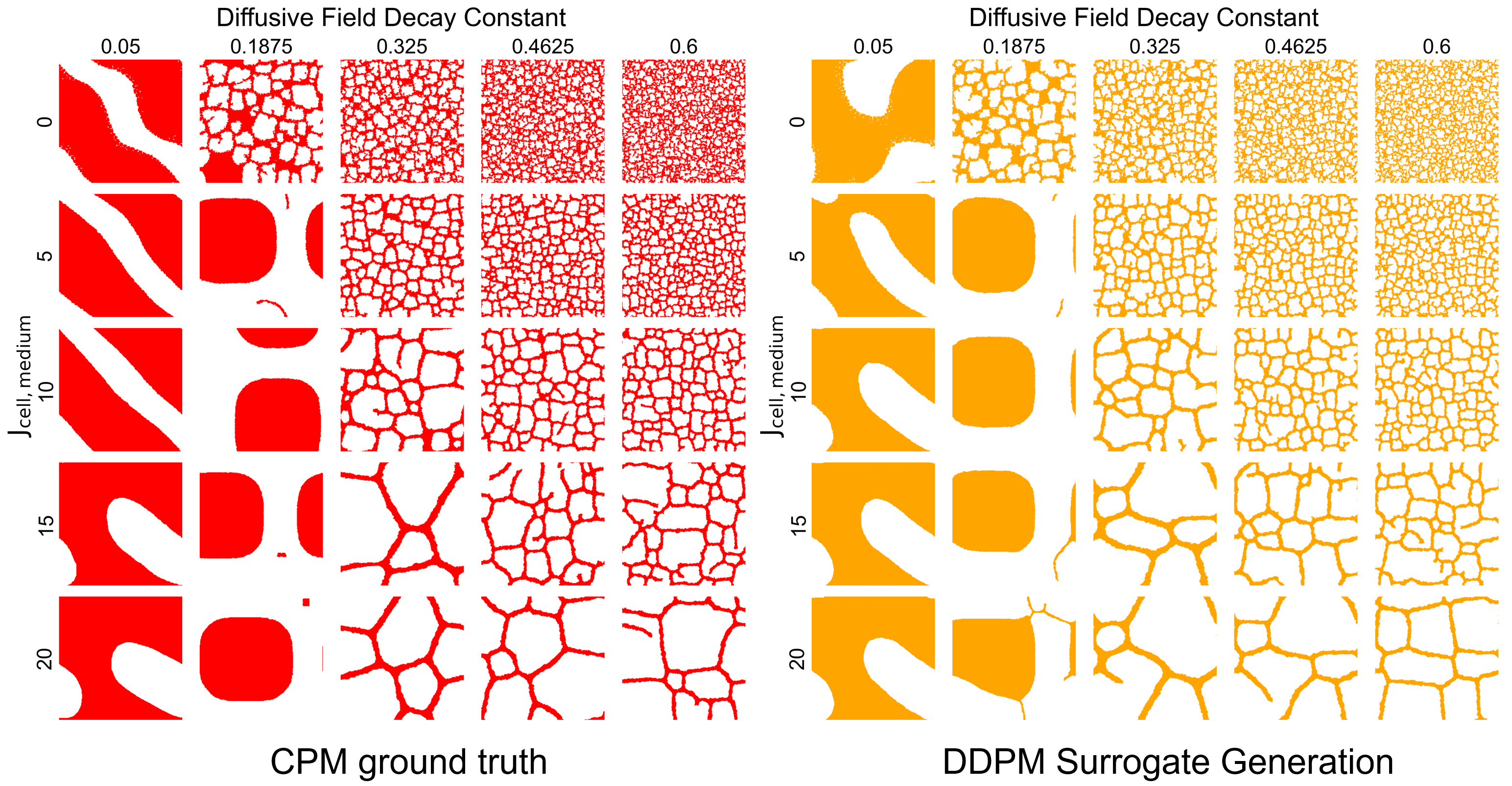}
    \caption{DDPM surrogate generates representative CPM configurations. The selected DDPM model surrogate generates visually representative CPM configurations throughout the parameter space of interest. Ground truth from representative CPM simulations at t = 20,000 MCS for all areas of the defined parameter space (left) is compared to surrogate model generated configurations (right).}
    \label{fig:comparison}
\end{figure}

After surrogate model selection, we generated a single CPM configuration for each class. This was performed without the added verification of the trained classifier to demonstrate the application of the DDPM alone as the CPM surrogate. The selected DDPM surrogate generated images for all areas of the parameter space and demonstrated generated configurations that visually resembled ground truth created directly from mechanistic CPM simulations (Figure \ref{fig:comparison}). The generated samples do appear imperfect in certain classes. Generations for diffusive field decay values of 0.05, contact energy $J$ values of 10, 15, and 20 (bottom 3 rows, first column) appear equivalent. These same three rows ($J$ values of 10, 15, 20) in the second column (diffusive field decay $k$ = 0.1875) also appear equivalent. While the latter three configurations do not resemble the representative samples of ground truth for these parameter sets, the patterns are present in the later frames of Movie 1, indicating that they are included in the training set.

\begin{figure}[htbp]
    \centering
    \includegraphics[width=0.7\textwidth]{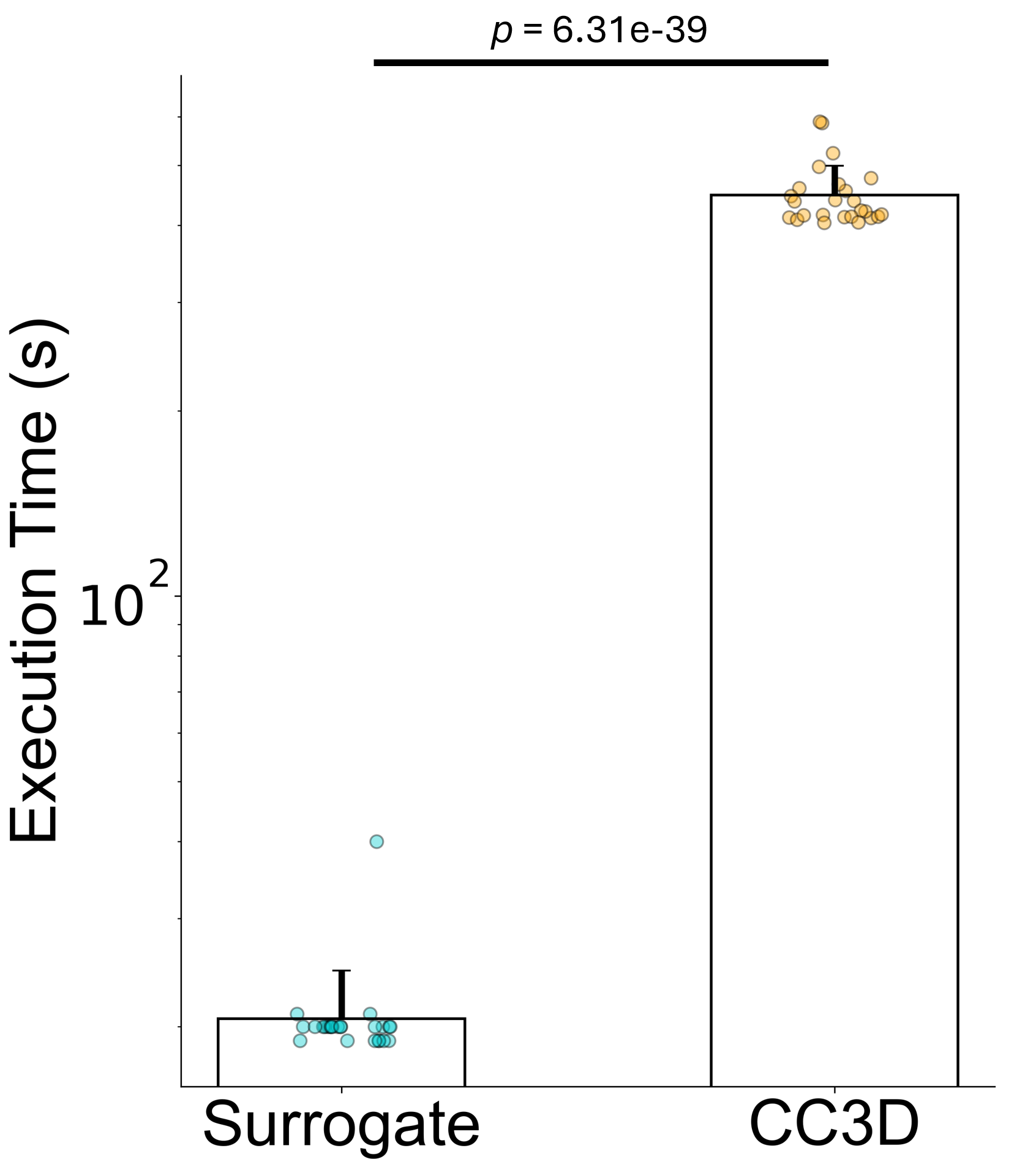}
    \caption{Surrogate model evaluates significantly faster than native CPM code execution. P=6.31e-39 by independent samples t-test for n=25 simulation configurations each. Mean DDPM surrogate model evaluations were 20.6s (std 4.0s) as compared with a mean of 447.6s (std 51.1s) for native CompuCell3D (CC3D) code execution. The surrogate model achieved approximately a 22x speed increase. Surrogate model evaluations were performed on a single node of 4 V100 GPUs, time recorded included loading the model into GPU memory and saving generated samples. CC3D simulation execution was performed on an HPC node consisting of AMD EPYC 9495 architecture.}
    \label{fig:speed}
\end{figure}

The trained surrogate model accelerates the evaluation of 20,000 MCS of the CPM when compared to simulation in native CompuCell3D (CC3D) software on our academic HPC cluster. Mean surrogate model evaluations were 20.6s (std 4.0s) as compared to a mean of 447.6s (std 51.1s) for CC3D code execution (Figure \ref{fig:speed}). Surrogate model evaluations were performed on a single HPC node consisting of 4 NVIDIA V100 GPUs and recorded time was inclusive of the time required to load the model into GPU memory and save the generated image. CC3D code was executed on an HPC node consisting of AMD EPYC 9495 processors.

\section{Discussion}
In this work, we leverage denoising diffusion probabilistic models as a mechanistic agent-based biological model surrogate. Additionally, we describe the use of an image classifier model as verification step to assist in model selection. Our surrogate produces representative model configurations at long timescales of 20,000 MCS and for multiple parameter sets of interest. Our model achieves a 22x increase in evaluation speed thus accelerating evaluation of a CPM.

Neural network-based surrogates allow for use of the graphics processing unit (GPU) to accelerate computational speed of a given task. The Cellular-Potts models as implemented in the widely available open-source frameworks CompuCell3D \citep{swat2012multi}, Morpheus \citep{starruss2014morpheus}, Chaste \citep{osborne2017comparing}, and Artistoo \citep{wortel2021artistoo} to name a few, evaluate the Cellular-Potts algorithm on the central processing unit (CPU). Additionally, these frameworks must explicitly evaluate model configurations at each MCS which results in an increased computational requirement and prolonged simulation evaluation time. Investigation into GPU-accelerated implementations for the Cellular-Potts algorithm is an active field of research. Recent methods allow for parallel evaluation of the Cellular-Potts algorithm using the GPU which allows for significant speed increases for model evaluation \citep{sultan2023parallelized, tapia2011parallelizing}. However, these GPU-accelerated Cellular-Potts implementations are not compatible with the widely adopted open-source CPM frameworks and may have limited ability to accommodate multi-method simulation such as diffusive fields described by systems of PDEs. Because of this, we were unable to perform a hardware-fair timing comparison between the DDPM surrogate on the GPU, and the Cellular-Potts model implementation on the GPU. Given the multiple iterative denoising steps required for sample generation from a DDPM, CPU evaluation of the DDPM surrogate would be computationally costly and disadvantageous for the purposes of a model surrogate. DDPMs and other neural network-based surrogates may be an additional method for allowing for GPU acceleration of mechanistic agent-based models \citep{comlekoglu2025surrogate}. 

We have demonstrated the novel use of image classifiers as a verification technique to assist in trained DDPM model selection and to evaluate a model’s potential for performance. In contrast, the Frechet’s Inception Distance (FID) \citep{heusel2017gans} is a metric developed for evaluating the generative modeling performance of generative adversarial networks (GANs) and has become popular for the evaluation of DDPMs as well \citep{karras2023analyzing, karras2022elucidating, karras2024guiding}. This scalar value defines a similarity of two distributions of images which is useful for comparing generative models trained on similar datasets such as the widely used open-source natural image dataset ImageNet \citep{russakovsky2015imagenet}. However, a scalar value is ill-suited to confirm for a user whether an image or generated model configuration exists within a specific parameter space or class. A lower FID would increase the probability that a generated sample exists within the intended class, but it is not confirmation of whether a specific generated sample does exist within the intended class. The FID is a global similarity metric calculated on the scale of tens-of-thousands of image samples rather than a per-sample confirmation of similarity. An image classifier, however, can confirm this similarity on a per-sample basis. Additionally, the FID may be less effective for our use case as many areas of the CPM parameter space yield visually similar configurations that may be difficult to identify in isolation. A trained image classifier will be able to confirm the identity of a generated configuration.

At present, our surrogate generates representative configurations at long timescales in 25 discrete areas of the parameter space. Generating reference configurations between these discrete parameter sets would be useful, however between-class interpolation to achieve this off-grid generalization is a complicated problem. The model described in this work is a class-conditional DDPM and methods of effectively interpolating between class labels is an area of active research for DDPM models. Multiple recent publications propose methods to achieve such interpolation including training additional neural network architectures to add conditioning controls to DDPM generation \citep{zhang2023addingconditionalcontroltexttoimage}, interpolating between generated samples in the latent space of a latent DDPM \citep{wang2023interpolatingimagesdiffusionmodels}, or mixing score functions of conditional generations \citep{rahimi2025scoremiximprovingfacerecognition}. While these methods prove useful for controlling generation of natural images, controlled generation between classes is particularly complicated in our use-case as a mechanistic model surrogate. Generating between-class samples derived through methods that rely on the discretely sampled training data does not guarantee that the resulting between-class image is representative of the corresponding area of the parameter space created by the mechanistic model. Additional methods must be developed to validate that between-class samples generated by the DDPM map effectively to those of the original mechanistic model. This is a non-trivial research activity that will be the focus of ongoing and future work.

An additional limitation of our DDPM model was the inability to achieve a perfectly uniform distribution of generated samples in some hyperparameter configurations as made obvious by the classifier verification and KL divergence. This phenomenon is an example of mode collapse and is a common occurrence in generative models \citep{durall2020combating}. Mode collapse is the phenomenon when a generative model is unable to demonstrate the diversity of data present in the training set. This is apparent in Figure \ref{fig:collapse} for DDPM configurations that fail to generate samples in all classes, or uniformly distributed across classes, even though the training data was uniformly distributed. 

In addition to model selection, the trained image classifier could have been used in combination with the DDPM to ensure that all generations of the surrogate are of the intended class. To do this, the DDPM surrogate could be repeatedly sampled until the classifier confirms that a sample from the intended class was generated. This configuration would define a hybrid DDPM-classifier surrogate where the classifier component may “rescue” a DDPM that may be difficult to optimally train at the cost of computational time and efficiency during generation. A hybrid surrogate model using the classifier together with the DDPM would address the observed mode collapse. Other proposed methods to alleviate mode collapse for generative diffusion models include elaborate architectures such as Diffusion-GAN hybrid models \citep{jiangzhou2024dgrm} or creative regularization and reinforcement learning-based fine-tuning strategies \citep{barcelo2024avoiding}. Diffusion models may also be controlled to produce outputs of interest via model guidance. Classifier guidance has been popularized by Dhariwal and Nichol \citep{dhariwal2021diffusion} whereby gradients of a pre-trained classifier steer sample generation from a DDPM towards a particular class or desired output. More recently, a method of classifier-free guidance has been described that requires the combination of a trained conditional and unconditional diffusion model to control generative sample diversity \citep{ho2022classifier}. Finally, a method of diffusion model autoguidance has been described that allows for the control and improvement of a DDPM generative output using a less-trained version of itself \citep{karras2024guiding}. Future efforts will investigate the potential for these techniques to improve upon a DDPM surrogate for a CPM.

At long timescales of 20,000 MCS, memory of an initial or reference configuration is lost. However, certain characteristics of the configuration such as the number and approximate size of circular vascular lacunae, the width of vascular branches, and the number of vascular sprouts are all functions of the underlying mechanistic model parameter values. These characteristic metrics define the phenotype or state of the biological system at a given set of parameters and were investigated in the original publication that described the CPM \citep{merks2006cell}. Additionally, the stochastic nature of these CPMs give rise to differences in model configurations between two simulation replicates with identical underlying parameters and initial configurations. Therefore, the characteristics of the class of configurations that arise due to a given set of parameters may be of primary importance to the investigation and application of stochastic agent-based mechanistic biological models. The image classifier that we have defined in this work has potential to define these characteristics and encode the \textit{universality class} of model configurations. This has significant implications for stochastic multicellular agent-based models as building upon this in future work may enable model interoperability between equivalent models encoded in different frameworks. At present, minor differences in algorithmic implementation of the for model methods such as the CPM result in drastic numerical differences for equivalent models encoded in different open-source frameworks of the same underlying mathematical method. This has been cited as a common and significant barrier to the interoperability, reusability, and reproducibility of stochastic multicellular models in systems biology \citep{niarakis2022addressing}.

While we use DDPMs in this work to model stochastic biological systems, these generative diffusion models are being adapted to study other stochastic physical systems. DDPMs and other generative model architectures have been used as model surrogates for the simulation of calorimeter showers to investigate fundamental particle experiments \citep{krause2024calochallenge, ahmad2024comprehensive}. Additionally, DDPMs are a key component in CorrDiff, cited as a component in a digital twin for weather forecasting \citep{mardani2023residual}. 

Our work demonstrates that a conditional DDPM architecture and its training procedure previously defined for generating natural images may be applied as a surrogate for Cellular-Potts models of biological systems. Our surrogate currently performs forward inference of 20,000 MCS across a discretely sampled parameter space. Looking ahead, an ideal digital twin would be capable of performing this inference at parameter sets between the demonstrated 25 classes, have some capability to solve for the inverse problem of parameter inference for mapping to its real-world biological counterpart, and allow for on-line updating. We previously discussed adapting data-driven between-class interpolation methods to provide a practical step towards delivering fine-grained forward inference across the parameter space\citep{zhang2023addingconditionalcontroltexttoimage, wang2023interpolatingimagesdiffusionmodels, rahimi2025scoremiximprovingfacerecognition}. For the inverse problem, we may consider adapting image-based parameter estimation methods that currently exist for mapping computational models to images of their biological counterparts\citep{hishinuma2025parameterestimation} as part of a greater digital twin infrastructure. For on-line updating, recent work leveraging diffusion models as aligned diffusion schrodinger bridges \citep{somnath2024aligneddiffusionschrodingerbridges} allows for learning the stochastic dynamics in between samples such that conditional generation of a resulting configuration given an arbitrary reference may occur. Thus, incorporating these existing advancements onto our work defines one potential path towards diffusion-model based digital twins.

\section*{Acknowledgements}
We acknowledge support from the following: Grant DOE ASCR \# DE-SC0023452 "FAIR Surrogate Benchmarks Supporting AI and Simulation Research" to Geoffrey Fox, NIH Grant GM131865 to Douglas W. DeSimone, James A. Glazier acknowledges partial support from National Science Foundation grants NSF 2303695 and NSF 2120200 and National Institutes of Health grant U24 EB028887. Tien Comlekoglu acknowledges support from National Institutes of Health grant T32-GM145443 and T32-GM007267. 

\section*{Data Availability}
Code for this work is publicly available at \url{https://github.com/tc2fh/CPM_DDPM_Surrogate}. Formatted dataset used for training the DDPM model is publicly available at \url{https://doi.org/10.5281/zenodo.17172845} \citep{comlekoglu_training_dataset}. Trained model checkpoints from Figure \ref{fig:collapse} are also included in the Zenodo dataset

\bibliographystyle{unsrtnat} 
\bibliography{references}

\clearpage
\section*{Supplementary Information}

\setcounter{figure}{0}
\renewcommand{\thefigure}{S\arabic{figure}}
\setcounter{table}{0}
\renewcommand{\thetable}{S\arabic{table}}

\begin{figure}[h!]
    \centering
    \includegraphics[width=\textwidth]{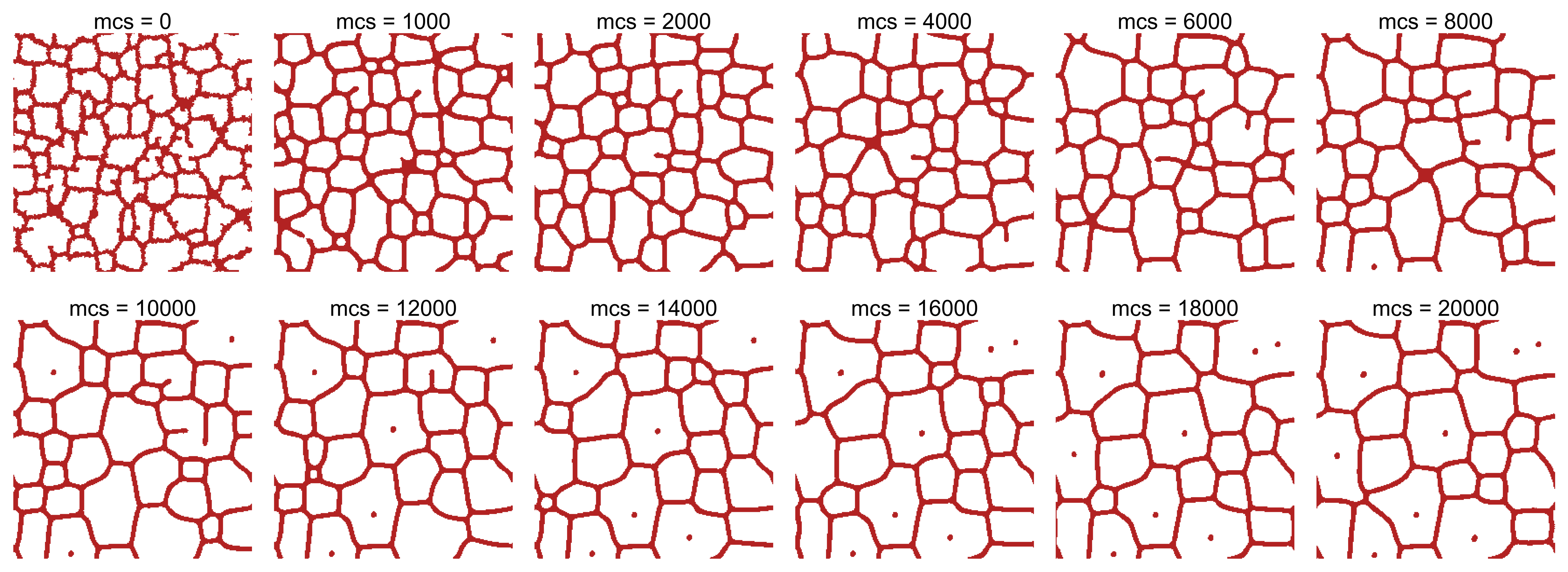}
    \caption{Prior deterministic neural-network based surrogate\citep{comlekoglu2025surrogate} fails at long time-horizons. Predicted model configuration visibly diverges from the reference phenotype (mcs = 0, initial condition, top-left) over long timescales of 20,000 MCS (bottom-right). Reference configuration and prior surrogate is defined for $J_{cell,medium} = 8.2, k = 0.6$}
    \label{fig:supps1}
\end{figure}

\begin{figure}[h!]
    \centering
    \includegraphics[width=\textwidth]{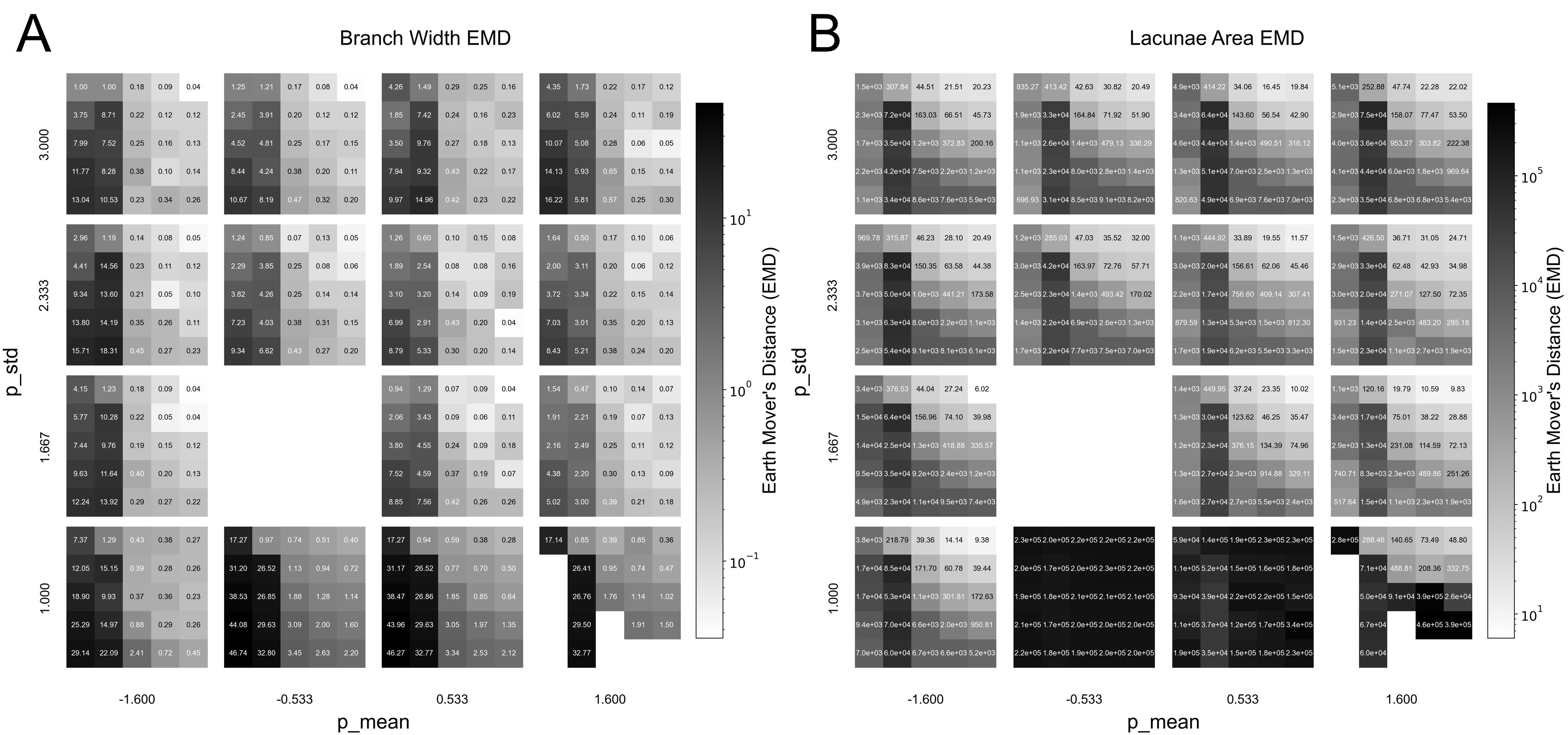}
    \caption{Feature-based distance metric comparisons of DDPM surrogate generations with training data. Earth Mover's Distance (EMD) calculated between class-conditional samples generated for Figure \ref{fig:collapse} and class-matched training samples. Calculated EMD values are displayed as a heatmap in the equivalent 5x5 grid layout shown previously in Figure \ref{fig:params}B, lower EMD values indicate greater similarity between distributions of features between surrogate-generated samples and CPM samples from the same class. (A) EMD from distributions of branch widths are displayed for each trained model. (B) EMD from distributions of lacunae area are displayed for each trained model. Large or absent EMD values are the result of generations that deviated sufficiently from the training data that these feature-based metrics could not be calculated.}
    \label{fig:supps2}
\end{figure}

\subsection{Feature-based metric calculations}
Our vascular networks may be described by the feature based metrics of branch width and lacunae area. To calculate a distribution of lacunae areas for a sample, we pattern each image above, below, to the left, and to the right to account for periodic boundary conditions. We label all connected lacunae regions to determine lacunae area in pixels, and drop duplicate regions identified through duplicate image inertia tensor eigenvalues for all identified lacunae regions. This inertia tensor calculation yields a mathematical representation of lacunae shape to allow us to conveniently identify duplicate regions. The distribution of areas for each sample was calculated for all DDPM generated samples. Distributions of branch widths were calculated by calculating the distance between the branch border and its skeletonized network representation for each voxel along the skeletonized network. All image processing operations to calculate distributions of vascular morphologic features used the open-source scikit-image library \citep{vanderwalt2014scikit} The Earth Mover's Distance (EMD) was calculated between the normalized feature distribution of DDPM generated samples and training data of the corresponding class.

\end{document}